\documentclass[12pt]{article}

\usepackage{graphicx}

\begin{document}
\begin{center}

{\bf A new model of arcsin-gravity }\\
\vspace{5mm}
 S. I. Kruglov
\footnote{serguei.krouglov@utoronto.ca}

\vspace{5mm}
\textit{Department of Chemical and Physical Sciences, University of Toronto,\\
3359 Mississauga Rd. North, Mississauga, Ontario, Canada L5L 1C6}
\end{center}

\begin{abstract}
The new model of modified $F(R)$ gravity theory with the function
$F(R) = R+(a/\gamma) \arcsin(\gamma R)$ is suggested and investigated.
Constant curvature solutions corresponding to the extremum of the effective potential are obtained. We consider both the Jordan and Einstein frames, and the potential and the mass of the scalar degree of freedom are found. It was shown that the de Sitter space-time is unstable but the flat space-time is stable. We calculate the slow-roll parameters $\epsilon$, $\eta$, and the $e$-fold number of the model. Critical points of autonomous equations for the de Sitter phase and the matter dominated epoch are obtained and learned.
\end{abstract}


\section{Introduction}

To describe the inflation and the present time universe acceleration one may try to modify the Einstein-Hilbert (EH) action of general relativity (GR). We analyze here the particular case of the $F(R)$ gravity model by replacing the Ricci scalar $R\rightarrow F(R)$ in EH action. It is known that $F(R)$ gravity models can describe the evolution of universe without introducing Dark Energy (DE) \cite{Appleby}, \cite{Capozziello}, \cite{Odintsov}. In such models the cosmic acceleration is due to modified gravity, and $F(R)$-gravity models may be an alternative to $\Lambda$-Cold Dark Matter ($\Lambda$CDM) model because of the new gravitational physics. In $\Lambda$CDM model the cosmological constant $\Lambda$ is introduced but there is a problem with the explanation of the smallness of $\Lambda$.

The first successful models of $F(R)$-gravity describing inflation were given in \cite{Hu}, \cite{Appl}, \cite{Star}, \cite{Odintsov1}, \cite{Odintsov2}, \cite{Odintsov3}, \cite{Starobinsky}.
Some $F(R)$-gravity models were discussed in \cite{Capozziello}, \cite{Odintsov}, \cite{Deser}, \cite{Kruglov}, \cite{Kruglov1}, \cite{Kruglov2}, \cite{Odintsov4}, \cite{Odintsov5}, \cite{Odintsov6} and in many other publications. It should be noted that $F(R)$-gravity is the phenomenological model that can describe the evolution of universe and gives the self-consistent inflation. It should be mentioned that models \cite{Odintsov1}, \cite{Odintsov2}, \cite{Odintsov3} represent the possibility to unify DE with inflation in a realistic way.

In this paper we introduce a new model of F(R) gravity that is a model for inflation and inflationary parameters
are studied.

In Sec.2, we formulate a model of $F(R)$ gravity with the dimensionless parameter $a$ and the parameter $\gamma$ with the dimension of (length)$^2$. We find the constant curvature solutions corresponding to the de Sitter space-time. In Sec.3, the potential and the mass of the scalar field are obtained in the Einstein (scalar-tensor) form of the model. The graphs of the functions $\phi(\gamma R)$, $V(\gamma R)$, and $m^2_{\phi}(\gamma R)$ are represented. It is shown that the de Sitter phase is unstable and the solution with the zero curvature scalar is stable. We calculate the cosmological parameters $\epsilon$, $\eta$, and the $e$-fold number of the model in Sec.4. In Sec.5 critical points of autonomous equations are studied. We estimate the function $m(r)$ characterizing the deviation from the $\Lambda$CDM model. The Sec.6 is devoted to a discussion.

The Minkowski metric $\eta_{\mu\nu}$=diag$(-1, 1, 1, 1)$ is used and $c$=$\hbar$=1 is assumed.

\section{The Model}

We consider the modified $F(R)$-gravity model with the function
\begin{equation}
F(R) = R+\frac{a}{\gamma}\arcsin(\gamma R),
\label{1}
\end{equation}
where $a$ is dimensionless parameter and $\gamma$ possesses the dimension of (length)$^2$.
The action of the model in the Jordan frame is given by
\begin{equation}
S=\frac{1}{2\kappa^2}\int d^4x\sqrt{-g}F(R),
\label{2}
\end{equation}
were $\kappa=M_{Pl}^{-1}$, $M_{Pl}$ is reduced Planck's mass,  and we do not include the matter Lagrangian density.
At $a=0$ one arrives at the EH action. According to Eq.(1) $F(0)=0$, so that $R=0$ corresponds to the flat space-time without cosmological constant. As GR passes local tests, we assume that the second term in Eq.(1) is small at low curvature regime, $\gamma R\ll 1$, $a<1$. We consider the description of the universe evolution in this model.

Classical stability takes place if $F'(R)>0$ \cite{Appleby} so that we require
\begin{equation}
F'(R)=\frac{a+\sqrt{1-(\gamma R)^2}}{\sqrt{1-(\gamma R)^2}}>0.
\label{3}
\end{equation}
Inequality (3) is satisfied for $a>0$, $\gamma R<1$. Quantum stability leads to inequality $F''(R)> 0$ \cite{Appleby}, which reads
\begin{equation}
F''(R)=\frac{a\gamma^2 R}{\left[1-(\gamma R)^2\right]^{3/2}}>0,
\label{4}
\end{equation}
and Eq.(4) is also satisfies at $a>0$, $\gamma R<1$.

\subsection{Constant Curvature Solutions}

Equations of motion \cite{Barrow}, in the case when the Ricci scalar $R_0$ is a constant, give the equation as follows:
\begin{equation}
2F(R_0)=R_0F'(R_0).
\label{5}
\end{equation}
It should be noted that constant curvature solutions of Eq.(5) correspond to the extremum of the effective potential. From Eqs.(1), (5) we find
\begin{equation}
2a\sqrt{1-(\gamma R_0)^2}\arcsin (\gamma R_0)=\gamma R_0\left(a-\sqrt{1-(\gamma R_0)^2}\right).
\label{6}
\end{equation}
The solution $R_0=0$ to Eq.(6) corresponds to the flat space-time. From Eq.(6) we obtain the parameter $a$ as a function of $x=\gamma R_0$:
\begin{equation}
a=\frac{x\sqrt{1-x^2}}{x-2\sqrt{1-x^2}\arcsin x}.
\label{7}
\end{equation}
The plot of the function $a(x)$ is given by Fig.1.
\begin{figure}[h]
\includegraphics[height=4.0in,width=4.0in]{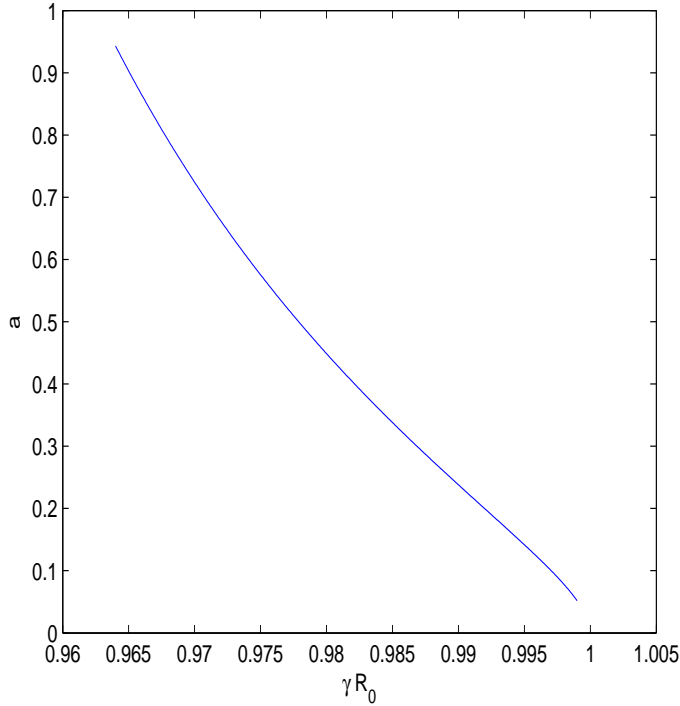}
\caption{\label{fig.1}The function $a$ versus $\gamma R_0$ corresponding to constant curvature solutions. }
\end{figure}
We obtain numerically solutions to Eq.(6) for different parameters $a$ which are given in Table 1.
\begin{table}[ht]
\caption{Constant curvature solutions}
\centering
\begin{tabular}{c c c c c c c c c c c c c c}\\[1ex]
\hline \hline 
$a$ & 0.1 & 0.2 & 0.3 & 0.4 & 0.5 & 0.6 & 0.7& 0.8 & 0.9 \\[0.5ex]
\hline 
x & 0.997 & 0.992 & 0.987& 0.982 & 0.978 & 0.974 & 0.971 & 0.968 & 0.965  \\
[1ex] 
\hline 
\end{tabular}
\label{table:Approx}
\end{table}
Solutions given in Table 1 correspond to the Schwarzschild-de Sitter space-time and the solution $R=0$ corresponds to the Minkowski space-time. The inequality $F'(R_0)/F''(R_0)>R_0$ \cite{Schmidt} assures that the constant curvature solutions describe DE which is future stable. This requirement leads to the equation
\begin{equation}
\left(1-x^2\right)\left(a+\sqrt{1-x^2}\right)>ax^2.
\label{8}
\end{equation}
The solution $R_0=0$ obeys Eq.(8) and, therefore, it is stable. The solutions given in Table 1 do not satisfy Eq.(8), and they lead to unstable de Sitter space-time. We will show that these solutions describe inflation and correspond to the maximum of the effective potential in the Einstein frame. As a result, even without matter, the model of Eq. (1) mimics DE and gives new non-trivial potential in the Einstein frame. The general study of de Sitter space-time solutions in $F(R)$ gravity models was done in \cite{Odintsov6}. Thus, we show that in our specific model based on Eq. (1) there are many solutions corresponding to the de Sitter space-time depending on the parameter $a$. This allows us to analyze the dependence of different universe scenarios on the parameter $a$.

\section{The Scalar-Tensor Formulation }

\begin{figure}[h]
\includegraphics[height=4.0in,width=4.0in]{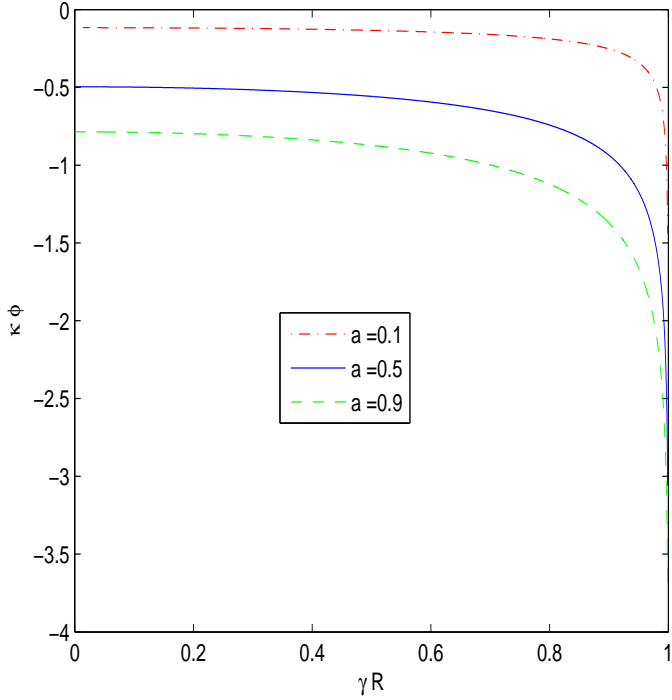}
\caption{\label{fig.2}The function  $\kappa\phi$ versus $\gamma R$. }
\end{figure}

\begin{figure}[h]
\includegraphics[height=4.0in,width=4.0in]{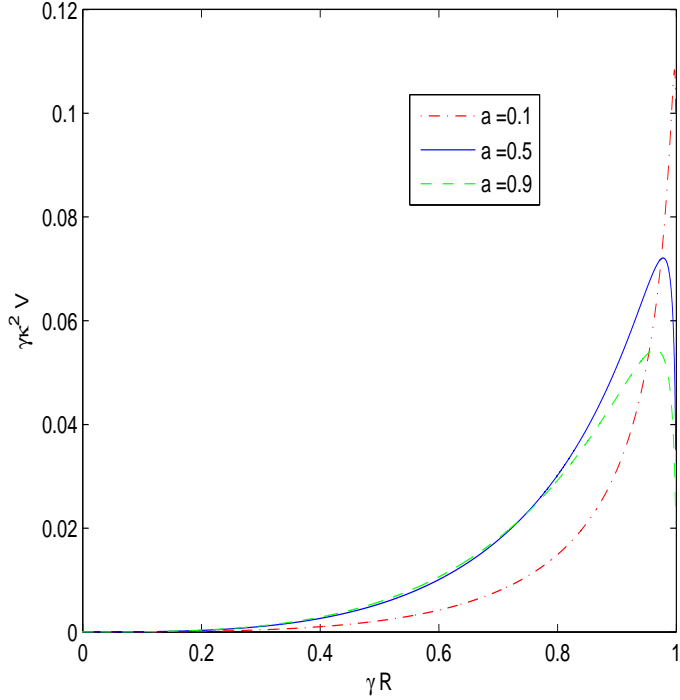}
\caption{\label{fig.3}The function $\gamma\kappa^2V$ versus $\gamma R$.}
\end{figure}

\begin{figure}[h]
\includegraphics[height=4.0in,width=4.0in]{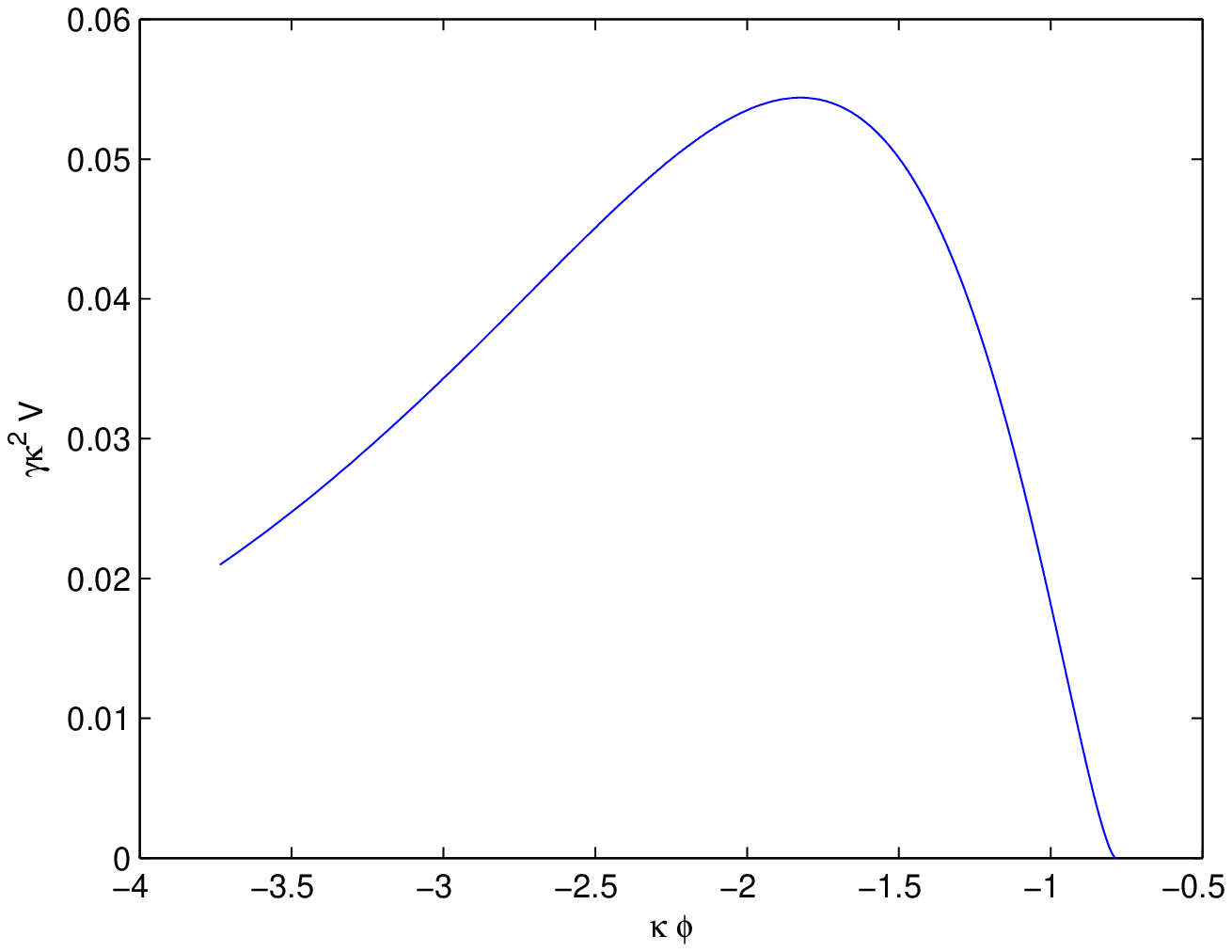}
\caption{\label{fig.4}The function $\gamma\kappa^2V$ versus $\kappa\phi$ ($a=0.9$).}
\end{figure}

One can investigate the theory in the Einstein frame to get the scalar-tensor form of the theory. For this purpose we perform the conformal transformation of the metric \cite{Sokolowski}
\begin{equation}
\widetilde{g}_{\mu\nu} =F'(R)g_{\mu\nu}=\frac{a+\sqrt{1-(\gamma R)^2}}{\sqrt{1-(\gamma R)^2}}g_{\mu\nu}.
\label{9}
\end{equation}
Then the Lagrangian density in Einstein's frame becomes
\begin{equation}
{\cal L}=\frac{1}{2\kappa^2}\widetilde{R}-\frac{1}{2}\widetilde{g}^{\mu\nu}
\nabla_\mu\phi\nabla_\nu\phi-V(\phi),
\label{10}
\end{equation}
where the Ricci scalar $\widetilde{R}$ in Einstein's frame is evaluated in new metric (9). The scalar field $\phi$ is given by
\begin{equation}
\phi(R)=-\frac{\sqrt{3}}{\sqrt{2}\kappa}\ln F'(R)=\frac{\sqrt{3}}{\sqrt{2}\kappa}\ln\frac{\sqrt{1-(\gamma R)^2}}{a+\sqrt{1-\left(\gamma R\right)^2}},
\label{11}
\end{equation}
and the potential $V(\phi)$ is
\[
V(R)=\frac{RF'(R)-F(R)}{2\kappa^2F'^2(R)}
\]
\vspace{-7mm}
\begin{equation}
\label{12}
\end{equation}
\vspace{-7mm}
\[
=\frac{a\sqrt{1-(\gamma R)^2}\left[\gamma R-\sqrt{1-(\gamma R)^2}\arcsin (\gamma R)\right]}{2\gamma\kappa^2\left(a+\sqrt{1-(\gamma R)^2}\right)^2}.
\]
The plots of the functions $\kappa\phi$ and $\gamma\kappa^2V$ versus $\gamma R$ are given in Fig.\ref{fig.2} and Fig.\ref{fig.3}. The plot of function $\gamma\kappa^2V$ versus $\kappa\phi$ for the value $a=0.9$ is represented by Fig.\ref{fig.4}.
It follows from Eq.(12) that the condition for an extremum of the potential $dV/dR=0$ leads to Eq.(5).
Therefore, the constant curvature solutions to Eq.(6) correspond to the extremum of the potential. Thus, the potential function (12) has the minimum at $R=0$ and the maximum given by Table 1. The flat (Minkowski) space-time ($R=0$) is the stable state and the states with the curvatures given in Table 1 are unstable.

From Eq.(12), we obtain the mass squared of a scalar degree of freedom (scalaron)
\[
m_\phi^2=\frac{d^2V}{d\phi^2} =\frac{1}{3}\left(\frac{1}{F''(R)}+\frac{R}{F'(R)}-\frac{4F(R)}{F^{'2}(R)}\right)
\]
\begin{equation}
=\frac{\sqrt{1-x^2}}{3\gamma}\biggl[\frac{\left(1-x^2\right)^{3/2}+a}
{ax\left(a+\sqrt{1- x^2}\right)}
-\frac{4\sqrt{1- x^2}\left(x+a\arcsin x\right)}{\left(a+\sqrt{1- x^2}\right)^2}\biggr].
\label{13}
\end{equation}
The plot of the function $\gamma m_\phi^2$ versus $x=\gamma R$ is represented in Fig.5.
\begin{figure}[h]
\includegraphics[height=4.0in,width=4.0in]{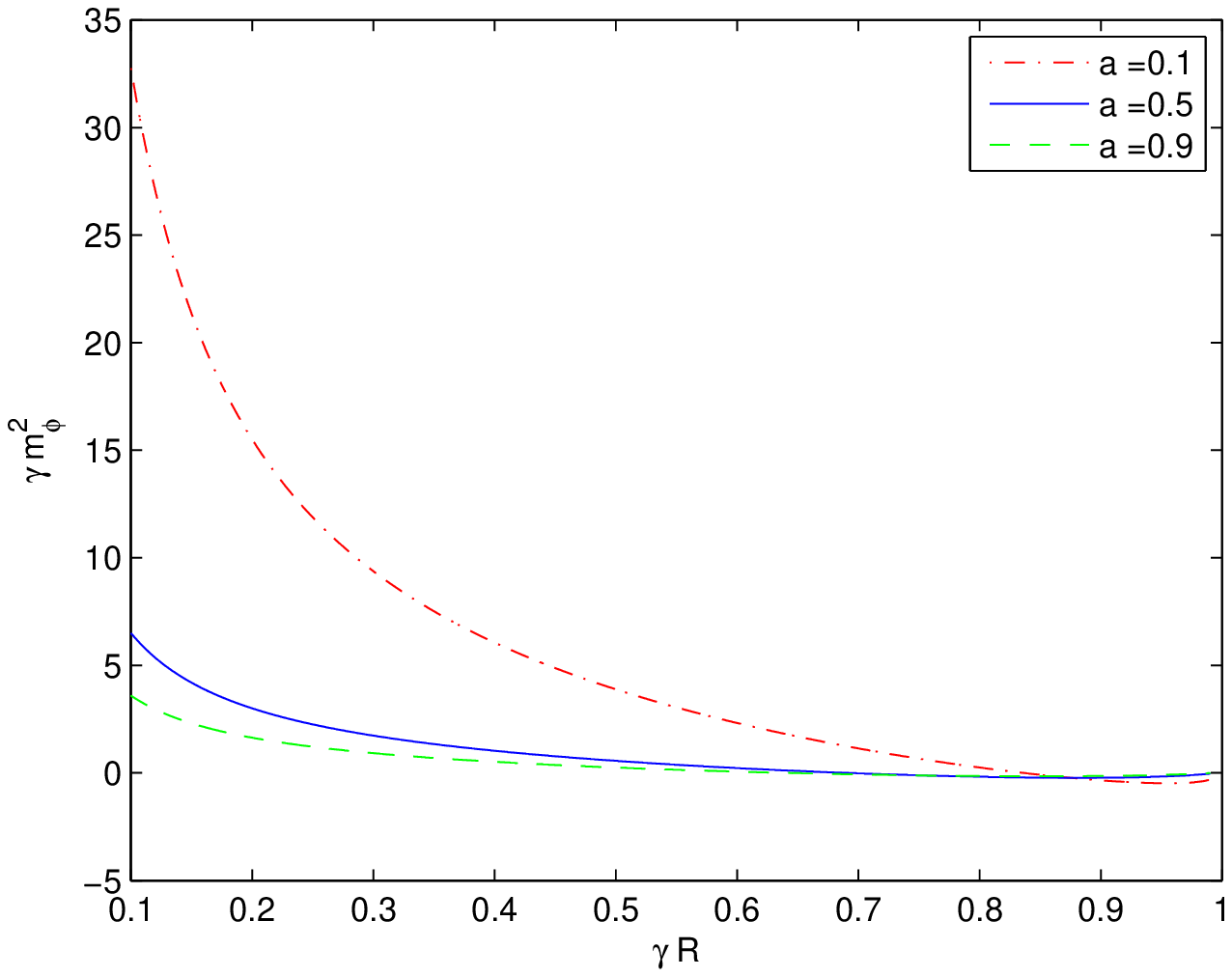}
\caption{\label{fig.5}The function $\gamma m^2_\phi$ versus $\gamma R$. }
\end{figure}
One can verify that $m^2_\phi<0$ for the constant curvature solutions given in Table 1, and, therefore, these solutions correspond to unstable states. The stability of the de Sitter solution in F(R) gravity models was first studied in \cite{Schmidt}. For small value of the parameter $\gamma$ the mass $m_\phi$ is big and, as a result, corrections to Newton's law are negligible.

Corrections of $F(R)$ gravity model will be small comparing to GR for $R\gg R_0$ ($R_0$ is a curvature at the present time) if the relations hold \cite{Appleby},
\begin{equation}
\mid F(R)-R\mid \ll R,~\mid F'(R)-1\mid \ll 1,~\mid RF''(R)\mid\ll 1.
\label{14}
\end{equation}
From Eqs.(3), (4) and (14) we obtain
\begin{equation}
\mid a\arcsin x\mid \ll x,~~~~ a \ll \sqrt{1-x^2},~~~~ ax^2\ll \left(1-x^2\right)^{3/2},
\label{15}
\end{equation}
where $x=\gamma R$. Eqs.(15) are satisfied if $0<a\leq 0.9$, $x< 0.43$.

\section{The Slow-Roll Cosmological Parameters}

Consider the slow-roll parameters \cite{Liddle}
\begin{equation}
\epsilon(\phi)=\frac{1}{2}M_{Pl}^2\left(\frac{V'(\phi)}{V(\phi)}\right)^2,~~~~\eta(\phi)=M_{Pl}^2\frac{V''(\phi)}{V(\phi)}.
\label{16}
\end{equation}
The slow-roll approximation occurs if conditions $|\eta(\phi)|\ll 1$, $\epsilon(\phi)\ll 1$ hold. When $\epsilon(\phi) < 1$ the acceleration of universe takes place and inflation ends if $\epsilon$, $|\eta|\approx 1$. At $|\eta(\phi)|< 1$, $\epsilon(\phi)< 1$ the kinetic energy of scalar field is small during inflation.
We obtain the slow-roll parameters from Eqs.(11)-(13),(16)
\[
\epsilon=\frac{1}{3}\left[\frac{RF'(R)-2F(R)}{RF'(R)-F(R)}\right]^2
\]
\begin{equation}
=\frac{1}{3}\left[\frac{x\left(\sqrt{1-x^2}-a\right)
+2a\sqrt{1-x^2}\arcsin x}{a\left(\sqrt{1-x^2}\arcsin x-x\right)}\right]^2,
\label{17}
\end{equation}
\[
\eta=\frac{2}{3}\left[\frac{F^{'2}(R)+F''(R)\left[RF'(R)-4F(R)\right]}{F''(R)\left[RF'(R)-F(R)\right]}\right]
\]
\begin{equation}
=\frac{2\left[a^2+\left(1-x^2\right)^2+a\sqrt{1-x^2}\left(2-5x^2-
4ax\arcsin x\right)\right]}{3xa^2\left(x-\sqrt{1-x^2}\arcsin x\right)},
\label{18}
\end{equation}
where $x=\gamma R$. The plots of the functions $\epsilon$, $\eta$ are represented in Figs.6 and 7.
\begin{figure}[h]
\includegraphics[height=4.0in,width=4.0in]{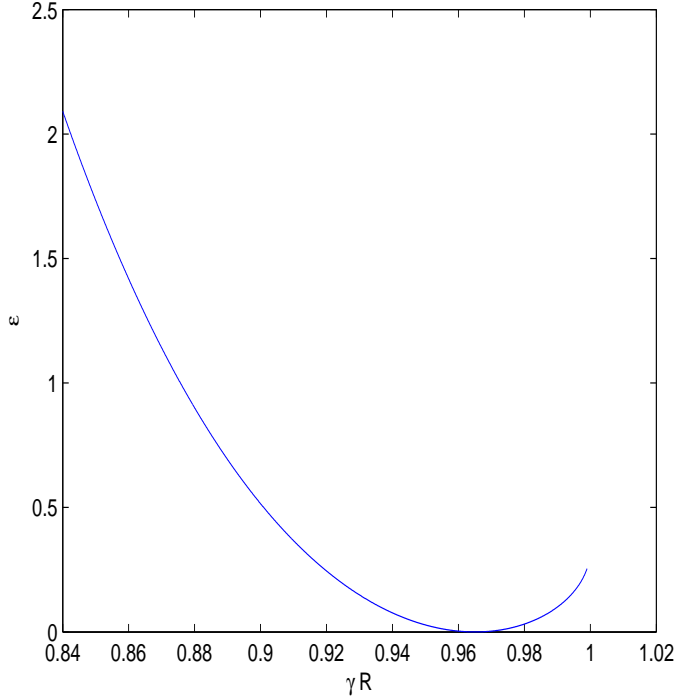}
\caption{\label{fig.6} The function $\epsilon$ versus $\gamma R$ (a=0.9).}
\end{figure}
\begin{figure}[h]
\includegraphics[height=4.0in,width=4.0in]{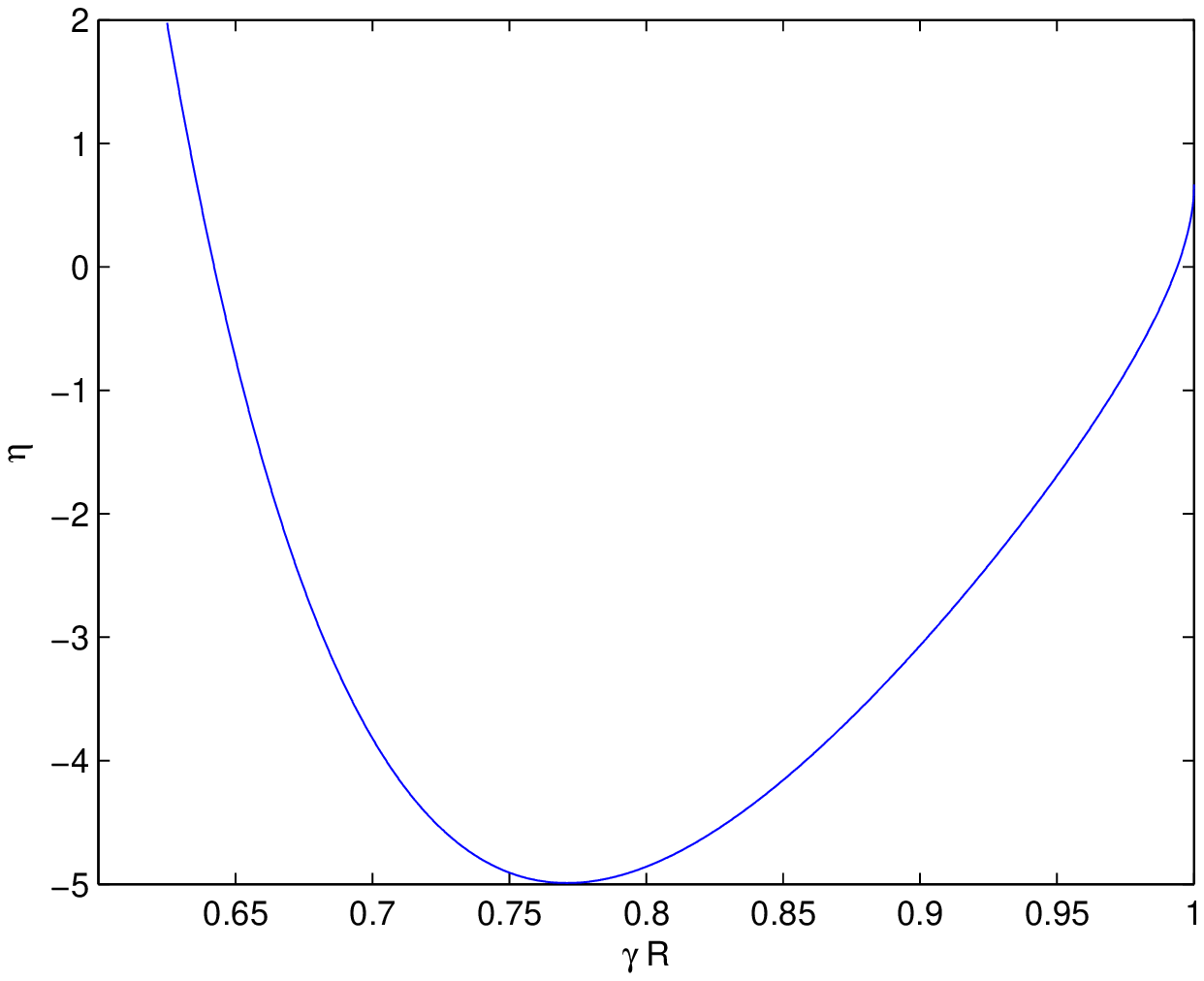}
\caption{\label{fig.7}The function $\eta$ versus $\gamma R$ (a=0.9). }
\end{figure}
The equation $\epsilon =1$ has the solution $x\approx 0.876$ for $a=0.9$ and $x\approx 0.892$ for $a=0.7$. It follows from Fig.6 that at $1>\gamma R>0.876$ ($a=0.9$) the inequality $\epsilon <1$ holds. The equation $|\eta | =1$ is satisfied at $x\approx 0.633$, $x\approx 0.653$ and $x\approx 0.971$ (see Fig.7). At $0.653>\gamma R>0.633$ and at $1>\gamma R>0.971$ ($a=0.9$), we have the result $|\eta | < 1$. Therefore, the slow-roll approximation, $\epsilon <1$ and $|\eta | < 1$, takes place at $1>\gamma R>0.971$ ($a=0.9$). The calculation of inflationary parameters allowed us to find the range of curvature when the slow-roll approximation of inflation takes place.

We can obtain the age of the inflation by calculating the $e$-fold number \cite{Liddle}
\begin{equation}
N_e\approx \frac{1}{M_{Pl}^2}\int_{\phi_{end}}^{\phi}\frac{V(\phi)}{V'(\phi)}d\phi.
\label{19}
\end{equation}
From Eqs.(11),(12) we find the number of $e$-foldings
\begin{equation}
N_e\approx \frac{3a^2}{2}\int_{x_{end}}^{x_0}\frac{x\left(x-\sqrt{1-x^2}\arcsin x\right)dx}
{(1-x^2)\left(a+\sqrt{1-x^2}\right)\left[\sqrt{1-x^2}\left(x+2a\arcsin x\right)-ax\right]},
\label{20}
\end{equation}
were $x_{end}=\gamma R_{end}$. It should be noted that integral (20) is singular at $x=\gamma R=1$ and, therefore, the value $N_e$ is sensitive for the range of integration.
We obtain the necessary amount of inflation $N_e\approx 40$ at $x_{end}=0.999$ and $x_0=0.95$ for $a=0.9$. Thus, the model under consideration describes the inflation and can give the reasonable age of the inflation.
The tensor-to-scalar ratio is given by the relation \cite{Liddle} $r=16\epsilon$. The PLANCK experiment gives the restrictions \cite{Ade} $r<0.11$. From this bound we obtain from Eq. (17) the range of curvatures $0.95769<\gamma R<0.97223$ ($a=0.9$) realizing the tensor-to-scalar ratio restriction. We note that $r= 0$ at $\gamma R\approx0.965$ ($a=0.9$).


\section{Critical Points of Autonomous Equations}

To investigate critical points of equations of motion, it is useful to introduce the dimensionless parameters \cite{Amendola} which become
\[
x_1=-\frac{\dot{F}'(R)}{HF'(R)}=-\frac{ax\dot{x}}{H\left(\sqrt{1-x^2}+a\right)\left(1-x^2\right)},
\]
\begin{equation}
~x_2=-\frac{F(R)}{6F'(R)H^2}=-\frac{\left(x+a\arcsin x\right)\sqrt{1-x^2}}{6\gamma H^2\left(a+\sqrt{1-x^2}\right)},
\label{21}
\end{equation}
\[
x_3=\frac{\dot{H}}{H^2}+2,
\]
\begin{equation}
m=\frac{RF''(R)}{F'(R)}=\frac{ax^2}{\left(1-x^2\right)\left(a+\sqrt{1-x^2}\right)},
\label{22}
\end{equation}
\[
r=-\frac{RF'(R)}{F(R)}=\frac{x_3}{x_2}=-\frac{x\left(a+\sqrt{1-x^2}\right)}{\sqrt{1-x^2}\left(x+a\arcsin x\right)},
\]
where $x=\gamma R$, $H$ is a Hubble parameter, and the dot means the derivative with respect to the time. The function $m(r)$ \cite{Amendola} shows the deviation from the $\Lambda$CDM model. Using (21), (22) equations of motion (in the absence of the radiation, $\rho_{\mbox{rad}}=0$) can be written in the form of autonomous equations \cite{Amendola}.
The critical points for the system of equations can be investigated by the study  of the function $m(r)$. The plot of the function $m(r)$ is given in Fig.8.
\begin{figure}[h]
\includegraphics[height=4.0in,width=4.0in]{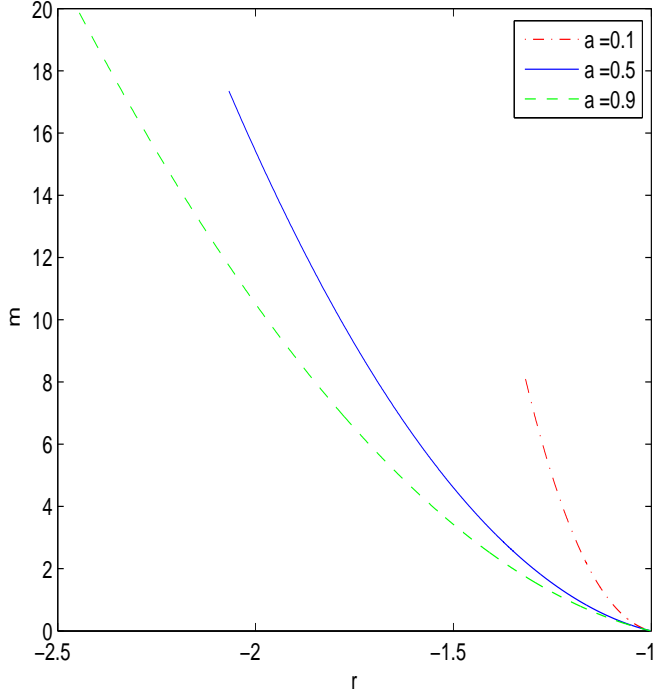}
\caption{\label{fig.8} The function $m(r)$}
\end{figure}
The de Sitter point $P_1$ \cite{Amendola} (in the absence of radiation, $x_4 = 0$) corresponds to the parameters $x_1=0$, $x_2=-1$, $x_3=2$ ($\dot{H}=0$, $r=-2$). One can check with the aid of Eqs.(6),(21) that $P_1$ corresponds to the constant curvature solutions. The parameter of matter energy fraction, $\Omega_{\mbox{m}}$, and the effective equation of state (EoS) parameter, $w_{\mbox{eff}}$, are
\begin{equation}
\Omega_{\mbox{m}}=1-x_1-x_2-x_3=0, ~~~~w_{\mbox{eff}}=-1-2\dot{H}/(3H^2)=-1.
\label{23}
\end{equation}
These parameters correspond to DE. The constant curvature solution $x \approx 0.965$ ($a=0.9$) gives unstable the de Sitter space because $1< m(r=-2)$ \cite{Amendola}.

For the critical point $P_5$ ($x_3=1/2$), $m\approx 0$, $r\approx -1$, and EoS of a matter era is $w_{\mbox{eff}}=0$ ($a=a_0t^{2/3}$). Then we have a viable matter dominated epoch prior to late-time acceleration \cite{Amendola}. The equation $m=-r-1$ has  the solution $m=0$, $r=-1$, $R=0$, corresponding to the point $P_5$.
One can verify with the help of Eq.(22) that $m'(r=-1)=0$ (see also Fig.8). Therefore the condition $m'(r=-1)>-1$ is satisfied and we have the standard matter era \cite{Amendola1}. As a result, there is the correct description of the standard matter era in the model under consideration.

\section{Conclusion}

The new model of modified $F(R)$ gravity is suggested and investigated. This is the effective gravity model which can describe the evolution of universe. The model gives the detailed account of the acceleration of universe corresponding to the de Sitter space and constant curvature solutions. It was shown that the de Sitter space-time is unstable because it corresponds to the maximum of the effective potential. We have considered both the Jordan and Einstein frames and obtained the potential and the mass of the scalaron. It was demonstrated that flat space-time is stable. The slow-roll parameters $\epsilon$, $\eta$ and the $e$-fold number of the model were calculated. We demonstrate that the model can give the reasonable age of the inflation. The analysis of critical points of autonomous equations shows that the standard matter era exists and the necessary conditions for the standard matter era are satisfied in the model considered. Thus, the model can be alternative to GR, and can describe early-time inflation.

\end{document}